\documentclass[prb,twocolumn,superscriptaddress,floatfix]{revtex4}
\usepackage{yfonts}
\usepackage{hyperref}
\usepackage{graphicx} 
\usepackage{amsmath}
\usepackage[dvips]{epsfig}
\newcommand{\beq}{\begin{equation}}
\newcommand{\eeq}{\end{equation}} 
\newcommand{\beqa}{\begin{eqnarray}}
\newcommand{\eeqa}{\end{eqnarray}}
\newcommand{\ba}{\begin{array}}
\newcommand{\ea}{\end{array}}

\begin{document}

\title{Rotating $^3$He droplets}

\author{Mart\'{\i} Pi}
\affiliation{Departament FQA, Facultat de F\'{\i}sica,
Universitat de Barcelona. Diagonal 645,
08028 Barcelona, Spain}
\affiliation{Institute of Nanoscience and Nanotechnology (IN2UB),
Universitat de Barcelona, Barcelona, Spain.}

\author{Francesco Ancilotto}
\affiliation{Dipartimento di Fisica e Astronomia ``Galileo Galilei''
and CNISM, Universit\`a di Padova, via Marzolo 8, 35122 Padova, Italy}
\affiliation{ CNR-IOM Democritos, via Bonomea, 265 - 34136 Trieste, Italy }

\author{Manuel Barranco}

\affiliation{Departament FQA, Facultat de F\'{\i}sica,
Universitat de Barcelona. Diagonal 645,
08028 Barcelona, Spain}
\affiliation{Institute of Nanoscience and Nanotechnology (IN2UB),
Universitat de Barcelona, Barcelona, Spain.}
\affiliation{Universit\'e Toulouse 3, Laboratoire des Collisions, Agr\'egats et R\'eactivit\'e,
IRSAMC, 118 route de Narbonne, F-31062 Toulouse Cedex 09, France
}

\begin{abstract} 
Motivated by recent experiments, 
we study normal-phase rotating $^3$He droplets  within Density Functional Theory in a semi-classical approach.
The sequence of rotating droplet shapes as a function of angular momentum 
are found to agree with those of rotating classical droplets,
evolving from axisymmetric oblate to triaxial prolate to two-lobed shapes
as the angular momentum of the droplet increases.
Our results, which are obtained for droplets of nanoscopic
size, are rescaled to the mesoscopic size characterizing ongoing
experimental measurements, allowing for a direct comparison of shapes.
The stability curve in the angular velocity-angular momentum plane
shows small deviations from the classical rotating drop model predictions,
whose magnitude increases with angular momentum.
We attribute these deviations to effects not included in the
simplified classical model description of a rotating fluid held together 
by surface tension, i.e. to surface diffuseness, curvature and finite compressibility,
and to quantum effects associated 
with deformation of the $^3$He Fermi surface.
The influence of all these effects is expected to diminish
as the droplet size increases, making the classical rotating droplet model
a quite accurate representation of $^3$He rotation.

\end{abstract} 
\date{\today}

%\pacs{05.30.Fk, 03.75.Ss, 67.85.-d}

\maketitle

\section{Introduction}

Helium is the only element in nature that may condense into macroscopic liquid samples at temperatures close to 
absolute zero. 
These systems can be made of either pure isotope, or of $^3$He-$^4$He isotopic mixtures. Below a temperature  $(T)$ that 
depends on the isotope (2.17 K for $^4$He, 2.7 mK for $^3$He) and also
 on the isotopic composition in the case of mixtures,\cite{Ebn70} these remarkable 
fluids undergo a well known normal-to-superfluid phase transition.
These properties, which are a manifestation at the microscale of the quantum nature 
of low temperature  liquid helium, have 
drawn a relentless  scientific  attention since they were uncovered about 80 years ago. 
A first hand, personal view of the first stages in the development of helium atoms 
beams and droplets has been given by J. Peter Toennies.\cite{Toe04a}

In more recent years, helium droplets have been the subject of renewed interest,
both experimentally and theoretically.
In the experiments, $^4$He droplets are produced at $T \sim$ 0.37 K;\cite{Toe04} hence, they are superfluid
and represent ideal ultra-cold matrices for spectroscopy studies of captured  
molecular impurities,\cite{Leh98} and for addressing superfluidity
at the  nanoscale.\cite{Toe04} 
At variance, $^3$He droplets are created at $T \sim$ 0.15 K;\cite{Sar12} hence, they are  
in the normal phase. For this reason, they are expected to behave more 
as classical viscous fluid droplets. $^3$He droplets have been the subject of far less
studies than $^4$He droplets, especially in recent times when the situation has been further aggravated by the
prohibitive price of this scarce helium isotope. The activity on $^3$He droplets has been 
partially reviewed in Refs. \onlinecite{Bar06,Anc17}.

Recently, large $^4$He droplets made  of $10^8-10^{11}$ atoms have been  
created by the hydrodynamic instability of a liquid helium 
jet passing through the nozzle of the molecular beam apparatus, as reviewed in Ref. \onlinecite{Tan18}. 
Helium drops, which are produced in the normal, non-superfluid
phase, may acquire angular momentum
during the passage of the fluid through the nozzle, before cooling down 
and become superfluid.
Such droplets could be analyzed one-by-one by x-ray and XUV
light and intense high harmonics sources\cite{Gom14,Lan18}  which have allowed to determine their shapes and,  doping 
them with Xe atoms,\cite{Gom14,Ges19} the presence of vortices, thus stressing their superfluid
nature. 
 
One of the intriguing findings of Refs. \onlinecite{Gom14,Lan18} was that the sequence of shapes of
the spinning superfluid $^4$He droplets is in accordance 
with that of classical rotating droplets.\cite{Bro80,Hei06,But11,Bal15}
It has been shown that it is the presence of quantized vortices 
that confers to the superfluid droplets the appearance
of classical viscous droplets when they are set in rotation.\cite{Anc18} 

It is quite natural to ask  whether droplets made of liquid $^3$He do indeed rotate as classical
droplets made of normal fluid, and whether their
properties may be described by a more microscopic approach, instead of  that  successfully used for
viscous liquid droplets.\cite{Bro80,Hei06,But11,Bal15} 
Besides, a proper microscopic description of pristine  $^3$He droplets is a necessary step
towards the study of mixed $^3$He-$^4$He droplets, which represent the prototype of 
strongly correlated
Bose-Fermi liquid mixtures.
These are the goals of the present paper, that accompanies 
an experimental one on rotating large $^3$He droplets.\cite{Ver19} 

In this work we describe deformed droplets within a Density Functional Theory (DFT) formalism for liquid $^3$He.\cite{Str87} 
At the experimental temperatures, thermal effects  on the energetics and morphology of the droplet are negligible,\cite{Mat13a} so 
we shall use a  $T=0$ method. Zero temperature means here  a very low temperature, but above the $\sim 2.7$ mK at which 
$^3$He becomes superfluid.  

As $^3$He atoms are fermions, the DFT-Kohn-Sham (DFT-KS) approach should be the method of choice for this study.  
It has been used in the past to address spherical $^3$He droplets 
made of up to a few hundred atoms.\cite{Gar98} 
Deformed (doped) $^3$He droplets with 
a few tens of $^3$He atoms have been addressed as well within such approach.\cite{Mat13a} Let us mention that
Diffusion Monte Carlo calculations have been made for pure $^3$He$_N$ droplets up to $N=34$,\cite{Gua05} 
and exact diagonalization results exist for $^3$He$_4$ clusters doped with Cl$_2$.\cite{deL09}

Unfortunately,  the DFT-KS approach is unfeasible for large, deformed  $^3$He droplets as the ones investigated here.
The use of a DFT-KS scheme is unavoidable when shell effects are expected to play a role, as in small droplets;\cite{Mat13a,Bar06}
but for the experimental droplet sizes, of the order of $10^8-10^{10}$ atoms,\cite{Ver19} the shell structure cannot play any substantial role.
Besides, temperatures of the order of 100 mK have  been found to wash out the shell structure of mixed $^3$He-$^4$He 
droplets.\cite{Lea13}
Under these conditions, the use of a 
semiclassical approximation to the DFT formalism, as the one described in the following, is fully justified.
The finite viscosity of $^3$He at the experimental temperatures  adds further justification to  using 
classical or semiclassical methods to address rotating $^3$He droplets.

We have  thus resorted to a semiclassical approach, treating the $^3$He droplets in the DFT plus rotating Thomas-Fermi (TF) 
framework, which has been  successfully used in Nuclear Physics to address deformed nuclei.\cite{Gra78,Mun79,Gra81} 
The DFT-TF method is the only realistic framework that has the virtue of making numerical simulations affordable
and that can be extended in a natural way to mixed helium 
droplets at the experimental conditions.\cite{Bar06} 
 
This work is organized as follows. In Sec. II we present the 
DFT-TF method used to describe the $^3$He droplets. The results
are discussed in Sec. III, and a summary and outlook are given in Sec. IV. 
Details on the rotating TF approximation are given in the Appendix.
 
\section{Model}

Within DFT, the total energy $E$ of a $^3$He$_N$ droplet  at zero temperature is written as a functional ${\cal E}_c$ 
of  the $^3$He  atom density $\rho$, here taken from Ref. \onlinecite{Bar97}:
\begin{equation}
E[\rho] =  \int  d {\mathbf r}  \frac{\hbar^2}{2m^*} \tau  +  \int d{\mathbf r} \,{\cal E}_c[\rho]
\label{eq1}
\end{equation}
The first term is the kinetic energy of $^3$He with an effective mass $m^*$, and $\tau
$ is the kinetic energy density, both depending on $\rho$.
 In the  TF  approximation of Ref. \onlinecite{Bar97} (see also Ref. \onlinecite{Str87}),
\begin{equation}
\tau=  \frac{3}{5} (3\pi^2)^{2/3}  \rho^{5/3} +  \frac{1}{18} \frac{(\nabla \rho)^2}{\rho} 
\label{eq2}
\end{equation}

The second term in the previous equation is  
a Weizs\"acker-type  gradient correction 
which is necessary in order to have helium densities with an exponential fall-off 
at the surface.\cite{Lom73}
The energy functional Eq. (\ref{eq1})  together with the TF approximation Eq. (\ref{eq2}) have been found to accurately reproduce 
the equation of state of the homogeneous system and the correct value for the $^3$He surface tension \cite{Bar97}.

\begin{figure}[!]
\centerline{\includegraphics[width=1.0\linewidth,clip]{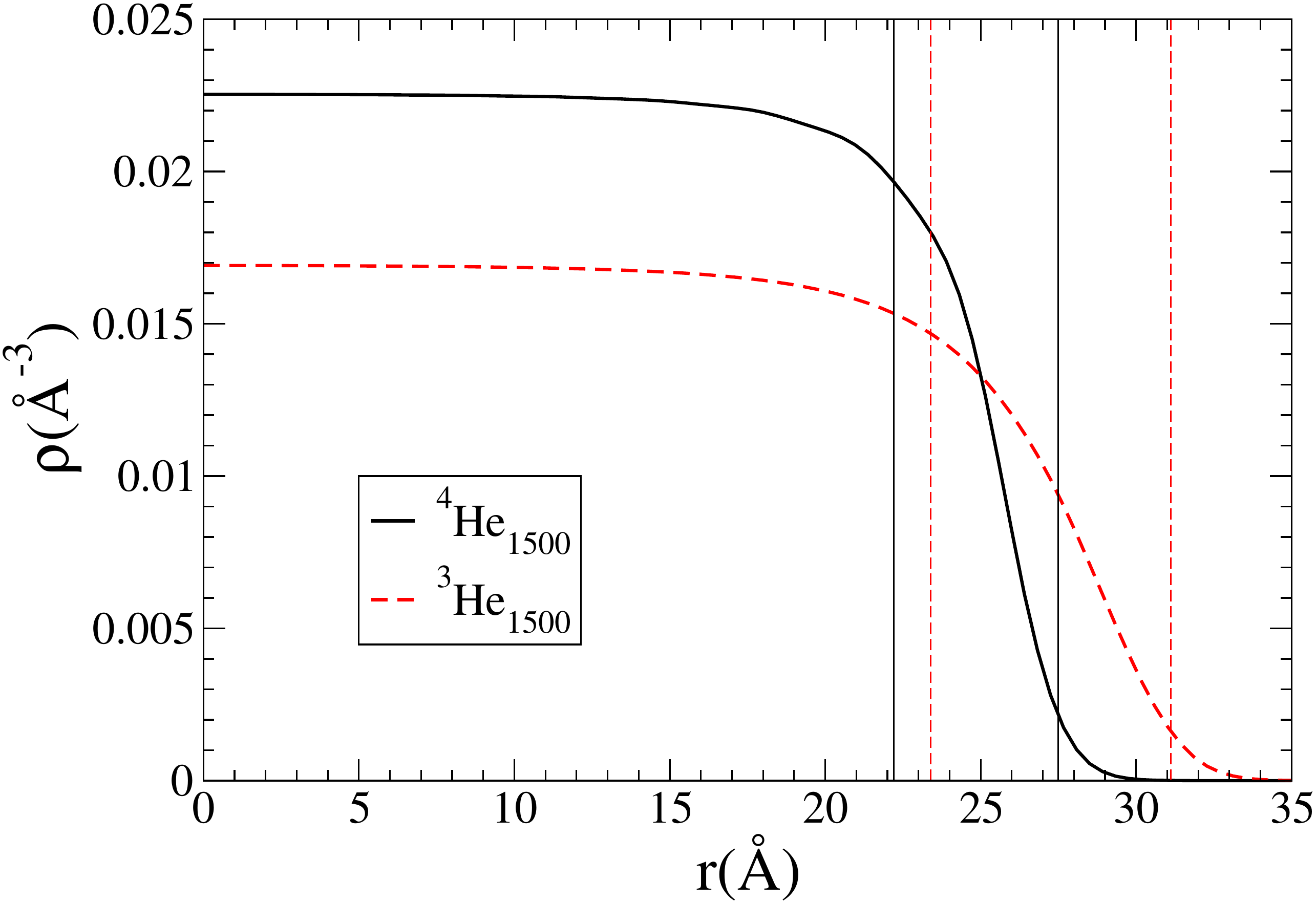}}
\caption{
Density profile of the $^3$He$_{1500}$ and $^4$He$_{1500}$ spherical droplets. The region between the thin vertical lines is the 
surface region defined as that where  
the helium density falls from  $0.9\times\rho_0$ to $0.1\times\rho_0$. 
}
\label{fig1}
\end{figure}

\begin{figure}[!]
\includegraphics[width=0.8\linewidth,clip]{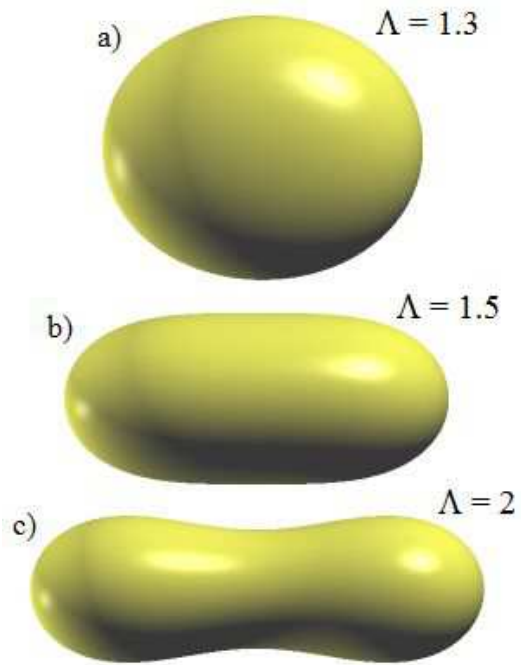}
\caption{
Some prolate $^3$He$_{1500}$ equilibrium configurations  represented by their sharp density surfaces (not to scale).
}
\label{fig2}
\end{figure}

\begin{figure}[!]
\centerline{\includegraphics[width=1.0\linewidth,clip]{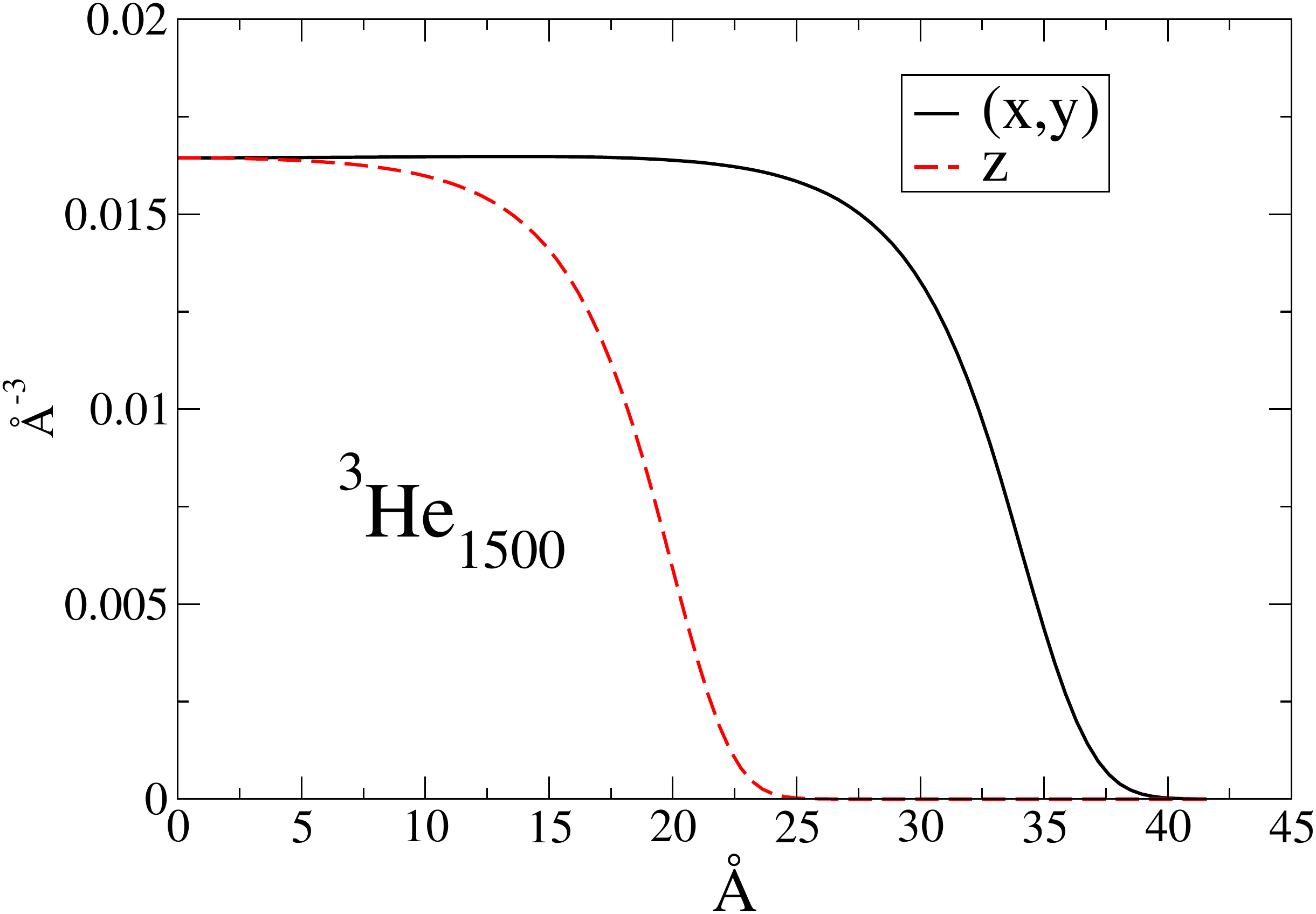}}
\caption{
Density profile of the oblate $\Omega=0.6364$ configuration along the  $x$ and $y$ axes (black, solid line), 
and along the rotation $z$-axis (red, dashed line). The densities have reflective symmetry 
with respect to the coordinate planes.
}
\label{fig3}
\end{figure}
 
 In this work, the number of $^3$He atoms is fixed to $N=1500$. 
 The droplet equilibrium configuration is obtained by solving the Euler-Lagrange (EL) equation arising from functional 
 variation of Eq. (\ref{eq1})
\begin{equation} 
\frac{\delta}{\delta \rho}
\left\{\frac{\hbar^2}{2m^*}\tau + {\cal E}_c \right\} = \mu 
\label{eq3}
\end{equation}
where $\mu$ is the $^3$He chemical potential. Defining $\Psi= \sqrt{\rho}$, Eq. (\ref{eq3}) can be written as a Schr\"odinger-like equation\cite{Bar97} 
\begin{equation}  
{\cal H}[\rho] \,\Psi  = \mu \Psi 
\label{eq4}
\end{equation}
where ${\cal H}$ is the one-body effective Hamiltonian that results from the functional variation.

When the rotating droplet --made of fermions in the normal phase-- is addressed 
in the TF approximation, the  Fermi sphere
is shifted by the motion of the  droplet as a whole; this adds to the droplet
total energy a rotational term that has the rigid body appearance\cite{Gra78}
\begin{equation}
E[\rho] \rightarrow  {\cal R} [\rho] = E[\rho] + \frac{1}{2} \,  {\cal I} \omega^2
\label{eq4b}
\end{equation} 
where  ${\cal R} [\rho]$ is the Routhian of the system and ${\cal I}$ is defined in Eq. (\ref{eq6})  below. Details are given in the Appendix.

To deposit  angular momentum in a droplet rotating with  angular velocity $\omega$ about a given axis (the $z$-axis here),
it is convenient to work in the  fixed-droplet frame of reference (corotating frame at  
angular velocity $\omega$), {\it i.e.} we consider 
\begin{equation}
E'[\rho] =   {\cal R} [\rho] - \hbar \omega \, \langle L \rangle  =
E[\rho] - \frac{1}{2}  \,  {\cal I} \, \omega^2
\label{eq5}
\end{equation}
where  
$\hbar \langle L \rangle=  {\cal I} \omega$ is the  $^3$He  angular momentum  obtained from the classical 
rigid body moment of  inertia  ${\cal I}$ 
\begin{equation}
{\cal I} = m\int d {\mathbf r} \,(x^2 + y^2) \rho({\mathbf r})
\label{eq6}
\end{equation}
We want to stress that the rigid-body moment of inertia is not an imposed ingredient to the DFT-TF framework. It arises naturally from the
TF approximation as shown in the Appendix. 
In  the more general DFT-KS framework,  the  rigid-body 
moment of inertia similarly appears within the   so-called
``cranking model'', as thoroughly discussed   for  nucleons  rotating in the mean field created by the atomic 
nucleus.\cite{deS74,Boh75,Rin80}

 \begin{figure}[!]
\centerline{\includegraphics[width=1.0\linewidth,clip]{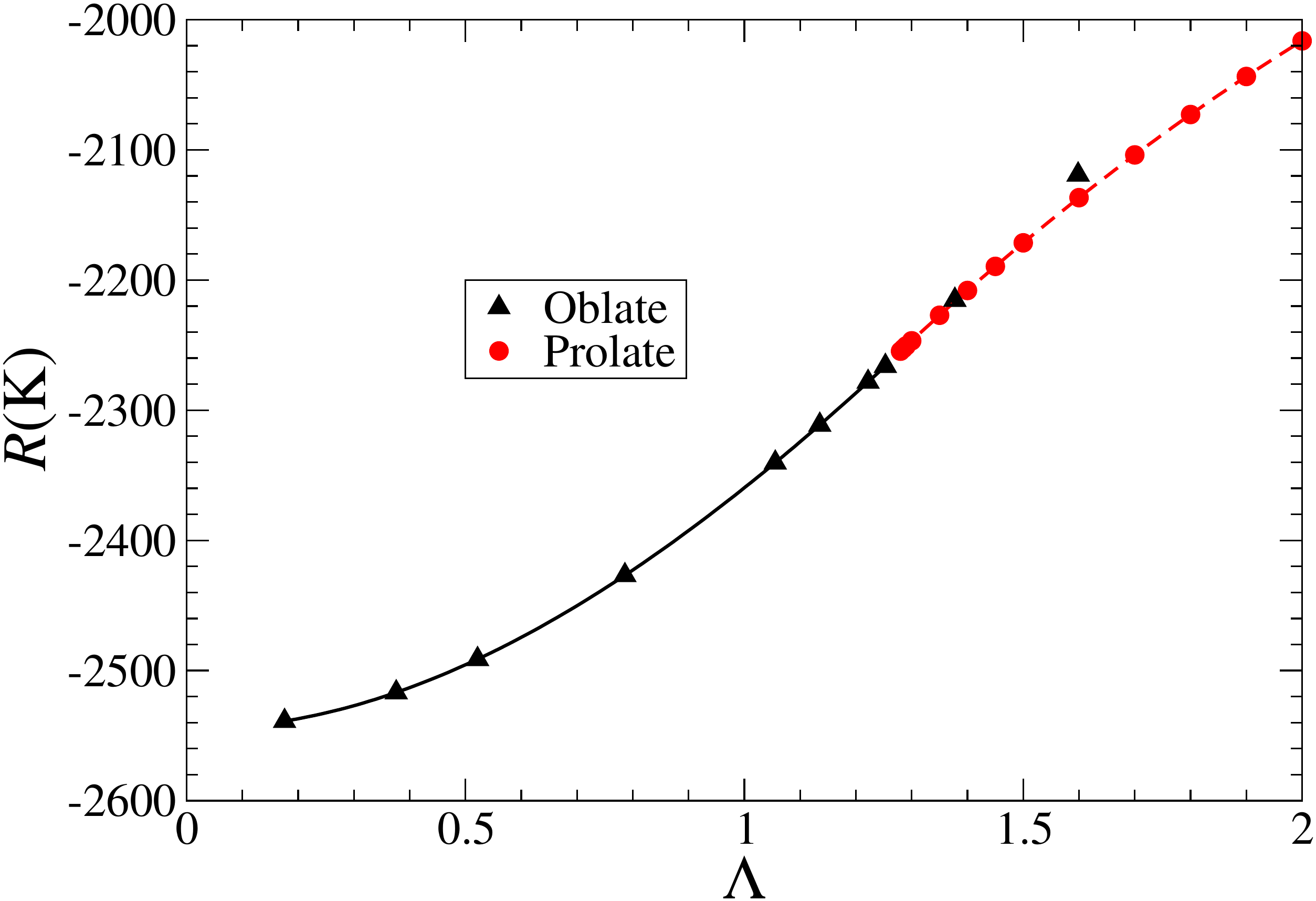}}
\caption{
 Routhian ${\cal R}[\rho]$ as a function of $\Lambda$.
Black triangles: oblate  configurations. Open triangles are metastable oblate configurations.  Red circles: prolate configurations.
The lines are cubic splines of the calculated points.
}
\label{fig4}
\end{figure}

 \begin{figure}[!]
\centerline{\includegraphics[width=1.0\linewidth,clip]{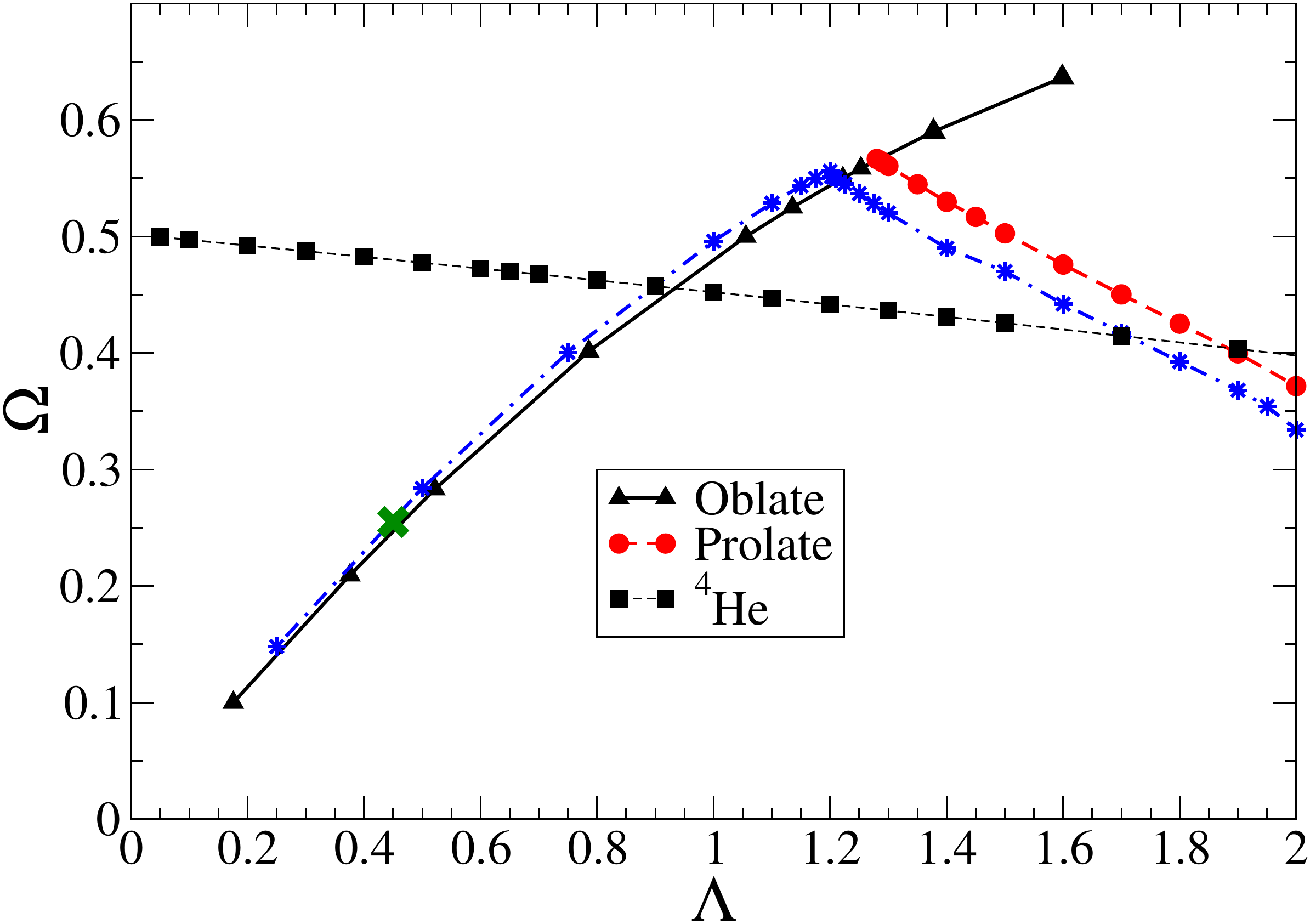}}
\caption{
Rescaled angular velocity $\Omega$ {\it vs.} rescaled angular momentum $\Lambda$.
Black triangles: oblate  configurations. Red circles: prolate configurations. Open triangles are metastable oblate configurations. 
The lines are  cubic splines of the displayed points. 
The  stars connected with a blue dot-dashed line is the  classical rotating drop result of Ref. \onlinecite{But11},
and the big green cross represents the average experimental value given in Fig. 5  of Ref. \onlinecite{Ver19}.
For the sake of comparison, the $\Omega(\Lambda)$ curve corresponding to vortex-free $^4$He droplets is also shown 
(black squares).
}
\label{fig5}
\end{figure}

In the corotating frame, we have to look for solutions of the EL equation resulting from the functional variation of  $E'[\rho]$: 
\begin{equation}
\left\{{\cal H}[\rho] \,- \frac{m}{2}\,\omega^2  (x^2+y^2)\right\} \,\Psi(\mathbf{r})  =  \,\mu \, \Psi(\mathbf{r})  \; .
\label{eq7}
\end{equation}
The results presented in this work have been obtained adapting  the  4He-DFT BCN-TLS computing package \cite{Pi17} 
to the case of $^3$He atoms in the TF approximation.
Details on how Eqs. (\ref{eq3}) and (\ref{eq7})  are solved can be found in  Refs. \onlinecite{Anc17,dft-guide} and references therein.
In short, we work in Cartesian coordinates, make extensive use of the Fast-Fourier Transform to compute the convolutions entering
the definition of the effective Hamiltonian  ${\cal H}$,  and obtain the droplet equilibrium configuration by imaginary-time relaxation.
To determine the prolate branch we have iterated on $\omega$ to get the desired 
$L_z$ value.\cite{Bro80,Hei06,But11,Anc18} 

The experimental droplets have  a radius in the 100-200 nm range.\cite{Ver19}
A comparison between the calculated nanoscopic DFT droplets and the experimental ones thus
requires some scaling procedure.
To this end, we have introduced a dimensionless
angular momentum $\Lambda$ and angular velocity $\Omega$  as done for classical drops\cite{Bro80,Hei06,But11}
\begin{eqnarray}
 \Omega &\equiv& \sqrt{\frac{m \, \rho_0 \, R^3}{8 \, \gamma}}\;  \omega 
 = \left[\frac{2 m}{\hbar^2} \frac{3}{64 \pi \gamma} N\right]^{1/2} \, \hbar \omega 
 \nonumber
 \\
\label{eq9}
\\
\Lambda &\equiv&\frac{\hbar}{\sqrt{8 \gamma R^7  m \rho_0}} \, L_z =
\left[\frac{ \pi }{3 \, \gamma \, r_0^4}\,\frac{\hbar^2}{2 m}\right]^{1/2}\, \frac{L_z}{N^{7/6}}
\nonumber
\end{eqnarray}
In the above  expressions,  $\gamma$  and $\rho_0$  are the surface 
tension and liquid atom density at zero temperature and pressure,
$R$ is the sharp radius of the  spherical  droplet when $L_z=0$, 
and $r_0$ is the bulk radius defined such that $4\pi r_0^3 \rho_0 /3= 1$, hence $R= r_0 N^{1/3}$. 
For liquid $^3$He, these values are  $\gamma$ = 0.113 K \AA$^{-2}$ and $\rho_0$ = 0.0163 \AA$^{-3}$
($r_0= 2.45$ \AA{}). Besides,  $\hbar^2/ m$ = 16.08 K \AA$^2$.
Liquid helium is fairly incompressible and hence the volume of the deformed configurations is also taken as 
$V= 4 \pi R^3/3$. 

In classical model approaches,\cite{Bro80,Hei06,But11}  
 it is assumed that the equilibrium configuration of the rotating droplet is solely determined by the balance between the 
 rotational and surface energies. This makes the problem  amenable to a  dimensionless 
formulation, 
where the Routhian can be expressed
 in terms of the    $\Lambda$ and $\Omega$ variables which characterize the equilibrium configuration
 irrespective of the droplet size and consequently the results are universal.  
This is quite not so in microscopic approaches such as DFT for instance,  where the droplets
have a large but finite  incompressibility, and also a surface finite width instead of a sharp
interface separating the fluid from the vacuum as assumed in classical models.
In general, not only surface and rotational energies matter to determine the shape of the droplet at equilibrium; volume and
quantum kinetic energy terms do change with deformation and this must be taken into account, as the  present DFT approach does.
Curvature energy, naturally incorporated in the DFT approach, contributes as well to the 
energy of the droplet, and its effect increases with the droplet deformation,
thus likely affecting the location 
of the higher angular momentum equilibrium configurations in the $\Lambda$-$\Omega$ plane. 

Consequently, some differences are expected to show up when comparing the results obtained in classical and DFT approaches, 
especially for small drops for which the surface thickness is not negligible compared to their radius. This is
shown in Fig. \ref{fig1} for the $^3$He$_{1500}$  and  $^4$He$_{1500}$ droplets as well. The $^3$He  surface
is thicker as a natural consequence of the the quantum zero point motion, which is larger for $^3$He than of $^4$He
because its mass is smaller. Experiments on the free surface of  liquid $^4$He films adsorbed on a solid substrate 
at $T=0.45$ K have yielded surface widths between $5.3 \pm 0.5$ \AA{}  (thin films) and 6.5 $\pm 0.5$ \AA{}  
(thick films).\cite{Pen00}
 
For any stationary configuration obtained solving 
Eq. (\ref{eq7}), a  sharp density surface is determined by calculating
the locus  at which  the helium density equals $\rho_0/2$; for a
spherical distribution this corresponds to a sphere of radius 
$R=r_0 N^{1/3}$. In the case of deformed droplets, three lengths  are introduced corresponding to
 the distances from the center  of mass (COM) of the droplet to the sharp surface along the rotation axis $(c_z)$,
the largest distance from the COM to the sharp surface along an axis perpendicular to the rotation axis $(a_x)$, 
and the distance of the COM to the sharp surface in the direction perpendicular to the other two $(b_y)$.
One expects,\cite{Lam45} and our calculations confirm, that $a_x \geq  b_y  > c_z$.

\begin{figure}[!]
\centerline{\includegraphics[width=1.0\linewidth,clip]{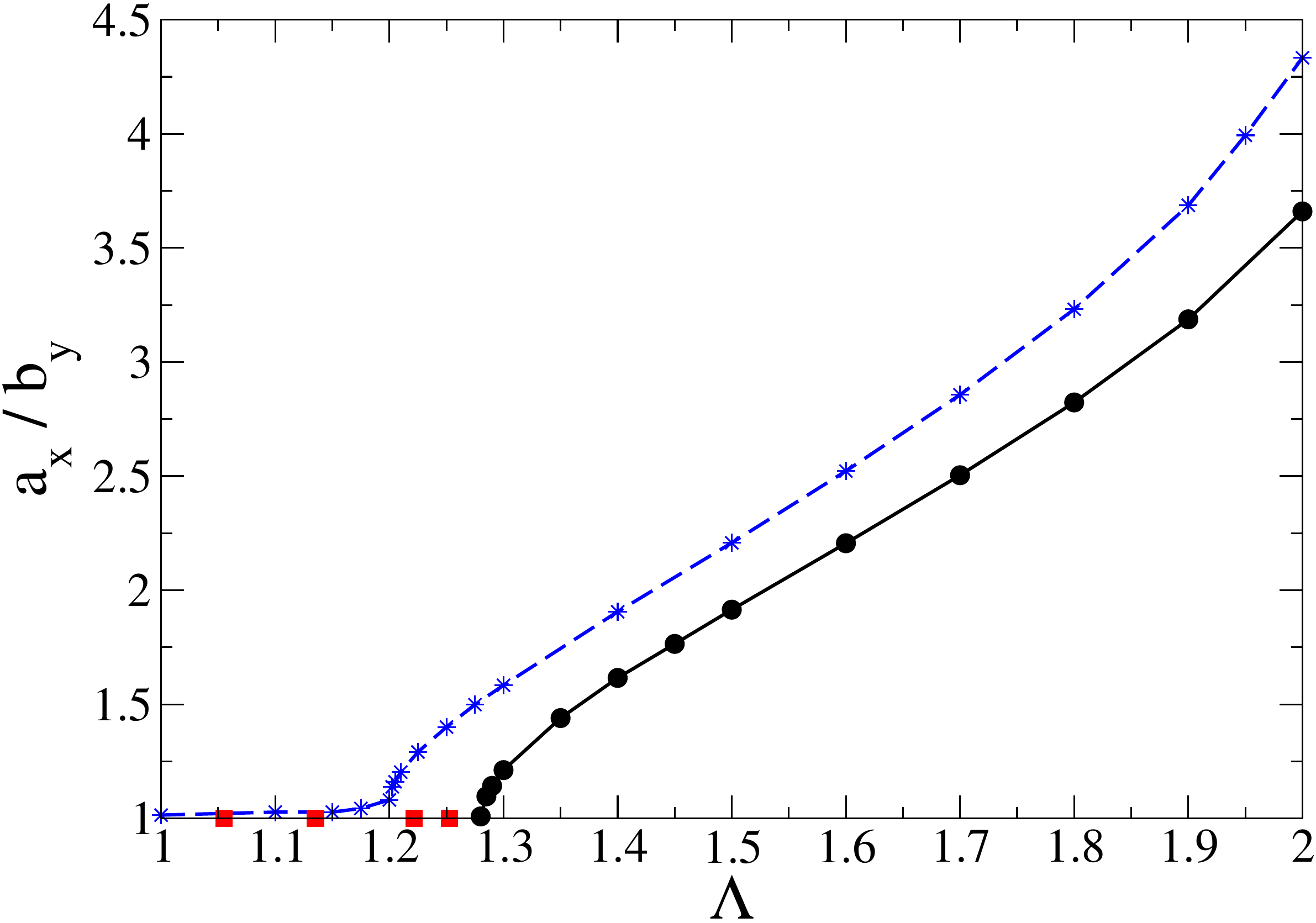}}
\caption{
Aspect-ratio   $AR=a_x/b_y$ curve {\it vs.} $\Lambda$ for $^3$He. $AR=1$ correspond to oblate configuration.
The starred symbols connected by a dashed blue line are the classical model results.\cite{Bal15}
}
\label{fig6}
\end{figure}
\begin{figure}[!]
\centerline{\includegraphics[width=1.0\linewidth,clip]{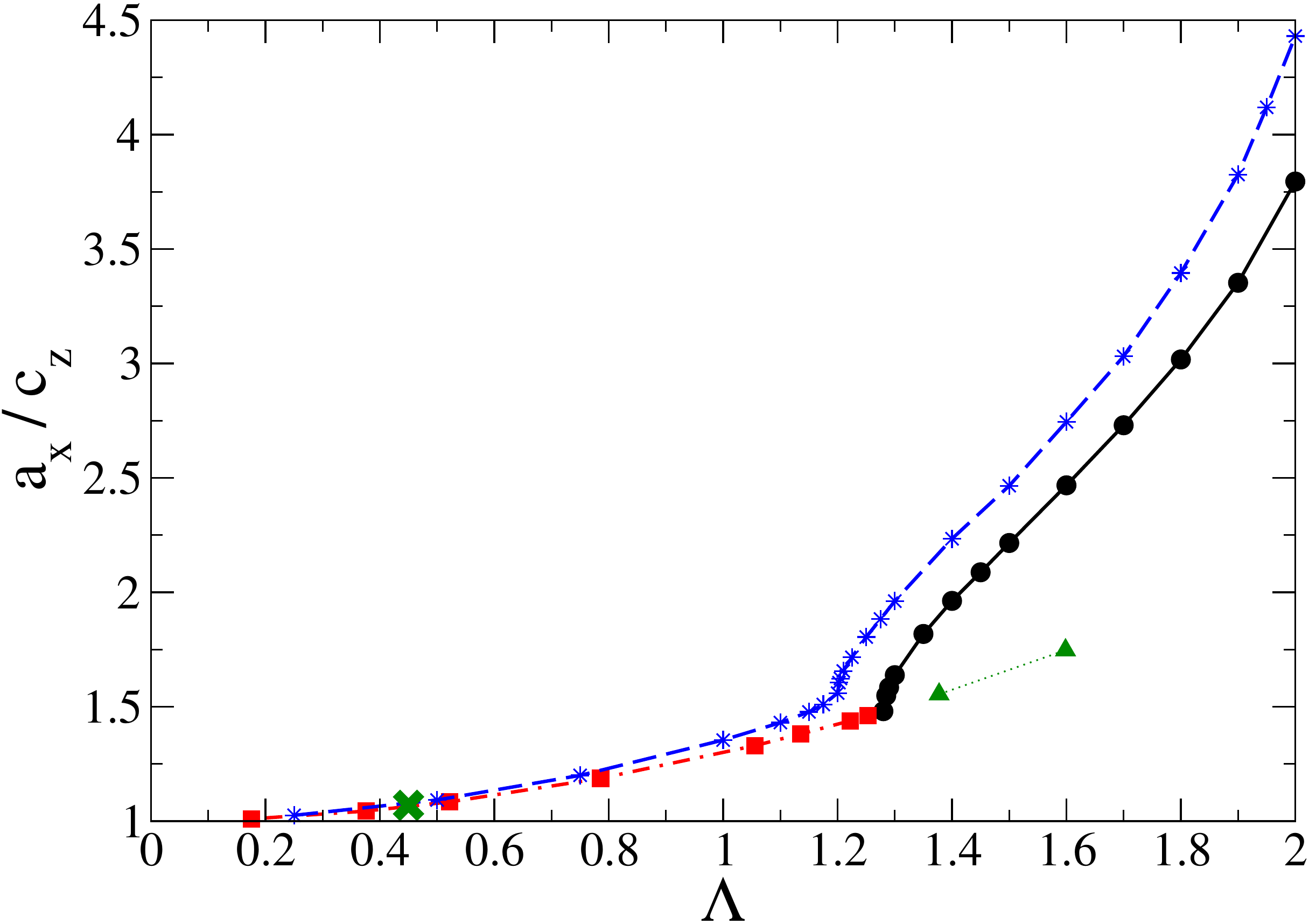}}
\caption{
Aspect-ratio   $a_x/c_z$ curve {\it vs.} $\Lambda$ for $^3$He. Red squares correspond to  oblate configurations,
and black dots to prolate configurations. Green triangles correspond to metastable oblate configurations.
The starred symbols connected by a dashed blue line are the classical model results,\cite{Bal15} 
and the big green cross represents the average experimental value given in Fig. 5  of Ref. \onlinecite{Ver19}.
}
\label{fig7}
\end{figure}
\begin{figure}[!]
\centerline{\includegraphics[width=1.0\linewidth,clip]{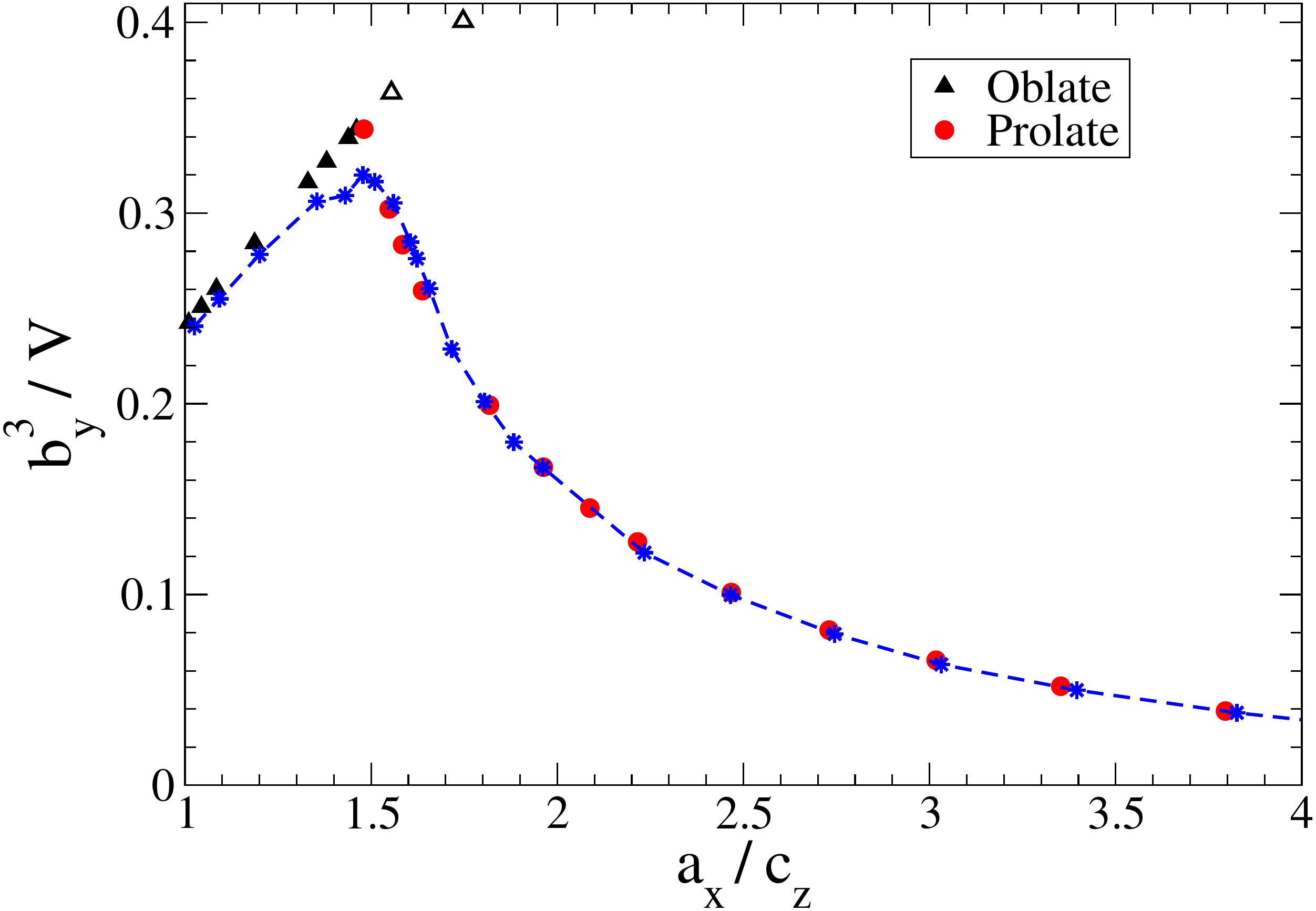}}
\caption{ 
Aspect-ratio   $b_y^3/V$ {\it vs.} $a_x/c_z$ curve  for $^3$He. Black triangles: oblate configurations. Open triangles are metastable
oblate configurations. Red circles: prolate configurations.
The starred symbols connected by a dashed blue line are the classical model results.\cite{Bal15}
}
\label{fig8}
\end{figure}

\section{Results}

\begin{table*}
%\begin{center}
\begin{tabular}{ccccccccccc}
\hline
\hline
& $\Lambda$ & $\Omega$ & $a_x$ (\AA)& $b_y$ (\AA) &  $c_z$ (\AA) & $AR$ & $b^3_y/V$ & $a_x/c_z$ 
& ${\cal I}/{\cal I}_{sph}$ & ${\cal R}$ (K)\\
\hline
\hline
O & 0.1755  & 0.1000   & 28.15  &  28.15  &   27.88  & 1 & 0.242    &  1.010   & 1.045  & -2538.86\\
O & 0.3759  & 0.2093   & 28.47   &  28.47   &   27.26  & 1 & 0.251    &  1.044   & 1.069  & -2516.80\\
O & 0.5216  & 0.2832   & 28.83   &  28.83   &   26.59 & 1  & 0.260    &   1.084  & 1.096  & -2491.30\\
O & 0.7858  & 0.4015   & 29.68   &  29.68   &   25.01 & 1  & 0.284    &  1.187   & 1.165  &-2426.83 \\
O & 1.0553  & 0.5000   & 30.76   & 30.76    &   23.12  & 1 & 0.316    &   1.330  & 1.256  &   -2340.33  \\
O & 1.1352  & 0.5250   &31.10    & 31.10    &   22.53 & 1  & 0.327  &   1.380  & 1.287   &    -2311.30      \\
O & 1.2217 & 0.5500   & 31.49    & 31.49     &   21.89 & 1 & 0.339   &  1.439    & 1.322  &      -2278.32   \\
O & 1.2527  & 0.5585  & 31.63    &  31.63    &  21.66  & 1 & 0.344    & 1.460    &  1.335  &   -2266.12   \\
O & 1.2800  & 0.5665   & 31.71  & 31.71  & 21.47  & 1 & 0.347  &  1.477  & 1.345  & -2254.65\\
O$^*$ &  1.3774 & 0.5900  & 32.21    & 32.21  &   20.72 & 1  & 0.363    &  1.555   & 1.389 &   -2215.34  \\                     
O$^*$ & 1.5984  & 0.6364  & 33.28   &  33.28   &  19.05  & 1  & 0.401    &  1.747  &   1.495   & -2119.10    \\
\hline
P  & 1.2800  & 0.5665  & 31.85    &  31.56   & 21.47  & 1.009  & 0.345  &  1.483   & 1.345   &   -2254.65   \\
P  & 1.2850  & 0.5652  & 33.21     & 30.27  &  21.43  & 1.097 & 0.301   & 1.550   & 1.353  &  -2252.65    \\
P  & 1.2900  & 0.5637  & 33.89     &  29.65  &   21.39  & 1.143 & 0.283  & 1.584    &  1.362 &  -2250.65  \\
P  & 1.3000  & 0.5605  & 34.89     &  28.79  &   21.30 & 1.212 & 0.259   &  1.638   & 1.380   &   -2246.66  \\
P  & 1.3500  & 0.5448  & 37.98     &  26.37  &   20.89 & 1.440 & 0.199   &  1.818   &  1.474  &   -2227.06 \\
P  & 1.4000  & 0.5296  & 40.17     & 24.85   &   20.47 & 1.616  & 0.167   &  1.963   & 1.573   & -2208.02 \\
P &  1.4500  & 0.5168  & 41.89     &  23.74  &   20.07 & 1.765  & 0.145   &  2.088   & 1.670 &  -2201.98   \\
P &  1.5000  & 0.5027  & 43.52    &  22.73   &   19.64  & 1.915 & 0.128    &  2.216 & 1.776  & -2171.43 \\
P &  1.6000  & 0.4759  & 46.40    &  21.03   &   18.80  & 2.206 & 0.101    &  2.467 & 2.001  & -2136.74 \\
P &  1.7000  & 0.4503  & 48.98    & 19.56    &   17.94 & 2.504  & 0.0813  &  2.730 & 2.247 & -2103.91 \\
P &  1.8000  & 0.4252  & 51.37    & 18.20    &   17.02 & 2.823  & 0.0655  & 3.018  & 2.519 & -2072.87 \\
P &  1.9000  & 0.3997  & 53.66    & 16.84    &   16.01 & 3.187  & 0.0519  & 3.352  & 2.829 & -2043.62 \\
P &   2.0000 & 0.3716  & 55.96     & 15.29   &  14.75 & 3.660  & 0.0389 & 3.795    & 3.203 & -2016.24 \\
\hline
\hline
\end{tabular}
%\end{center}
\caption{Characteristics of the  rotating $^3$He$_{1500}$ droplet configurations calculated in this work. 
O: oblate configurations; 
P: prolate configurations. O$^*$: metastable oblate configurations. $\Lambda$ and $\Omega$ are the dimensionless 
angular momentum and velocity, and ${\cal R}$ is the Routhian. $AR$ is the aspect ratio $AR=a_x/b_y$ ($AR=1$ for 
oblate configurations), and  ${\cal I}/I_{sph}$ is the DFT moment of inertia in units of 
that of a sphere of sharp radius, ${\cal I}_{sph} = (2/5) m \, r_0^2 \, N^{5/3}$; see the text for the meaning of the other  entries.
\label{Table1}
}
\end{table*}

Table \ref{Table1} collects the relevant features of the calculated stationary configurations.

Figure \ref{fig2} shows three characteristic prolate $^3$He$_{1500}$ 
droplets whose shapes evolve from ellipsoidal
to capsule-like to two-lobed as angular momentum increases. 
It is worth mentioning that vortex-free $^4$He droplet configurations are more stretched than $^3$He ones for the
the same $\Lambda$ value. In particular, the  $\Lambda=1.5$ $^4$He configuration is  already two-lobed and
has $a_x/c_z=3.578$, see the Supporting Material of Ref. \onlinecite{Anc18}.
This is due to the superfluid character of the $^4$He droplets. In the absence of vortex arrays, only capillary waves can carry angular 
momentum and this requires larger deformations than in the $^3$He case.

Blood-cell shapes (i.e. oblate droplets thinner in the center and 
thicker at the periphery) have been observed for 
spheroidal $^4$He droplets beyond the classical stability limit.\cite{Gom14} 
Metastable oblate $^3$He droplets display  a minute 
depletion at their center, as can be seen in Fig. \ref{fig3} for the 
largest angular velocity considered in our calculations, $\Omega=0.6364$.

To determine the oblate-to-prolate bifurcation point, one has to compare the  
Routhian  ${\cal R} [\rho]$ of the oblate and prolate configurations for the same  
$\Lambda$ value;  the configuration with the smaller ${\cal R}$ is the equilibrium one.\cite{Bro80}
Within the classical model approach to droplets subject only to surface tension and centrifugal forces,
the  bifurcation point is  
$(\Omega_{cl},\Lambda_{cl} )=$ (0.55,1.20).\cite{Bro80,Hei06,Bal15} 
Within the DFT-TF approach to $^3$He droplets, the bifurcation point  is at $(\Omega,\Lambda) \sim$ (0.57,1.28),
as can be seen from Fig. \ref{fig4}.

Figure \ref{fig5} shows the rescaled angular velocity $\Omega$ {\it vs.} rescaled angular momentum $\Lambda$. 
For  a fixed  $\Omega$, the DFT result is shifted  to the right of the classical one. 
The existence of a surface width is expected to produce some shift.
Indeed, due to the density spill-out beyond the sharp density surface, 
droplets described realistically have more fluid away from the rotation axis than if the surface is sharp as in classical models. 
Hence, for a given $\Omega$,  DFT configurations must have a larger
moment of inertia and thus  a larger  $\Lambda$ value.\cite{note}

Figure \ref{fig5} also shows the $\Omega(\Lambda)$ relationship 
for vortex-free $^4$He droplets --where angular momentum is
associated to giant capillary waves-- obtained in Ref. \onlinecite{Anc18}.  It is
worth seeing the completely different behavior between a rotational  ($^3$He) and an irrotational --potential-- fluid 
(superfluid $^4$He). 

The cross in the oblate branch in Figs. \ref{fig5} and \ref{fig7} shows the  experimental average
value measured for  $10^8-10^{11}$ atoms $^3$He droplets,\cite{Ver19} indeed
confirming the predictions of our calculations and the validity of the 
classical rigid-body model for $^3$He rotation.

Figure \ref{fig6} shows $AR$ as a function of $\Lambda$ extracted from the information in Table \ref{Table1}. For comparison, the
classical result is also shown.\cite{But11} 
Figure \ref{fig7} shows the aspect ratio $a_x/c_z$ as a function of $\Lambda$. 

We recall that
the diffraction images that are observed in experiments on spinning $^3$He droplets
do not allow to obtain the droplet image in the direction 
perpendicular to the detector plane.\cite{Ver19} Angular momentum is also a key quantity that has eluded {\it direct}
experimental determination for $^4$He\cite{Gom14,Lan18} and $^3$He droplets as well.\cite{Ver19} 
When the diffraction images correspond to droplets that  have been unambiguosly detected with their rotation axis aligned with 
the x-ray beam,\cite{Gom14,Ber17} the calculated 
aspect ratio  $AR=a_x/b_y$ as a function of $\Lambda$ might allow to determine the angular momentum of the droplet using 
the classical or DFT-TF calculations.   
The fact that both approaches
sensibly yield the same $\Omega(\Lambda)$ relationship in the oblate
branch renders  this model-dependent procedure to determine the angular momentum and velocity of oblate
$^3$He droplets very reliable up to  fairly large $\Lambda$ values. 
 
Our results confirm that the diffraction images of
Fig. 2(b) and (c) of Ref. \onlinecite{Ver19} indeed represent  ellipsoidal and  capsule-like 
droplet shapes, respectively; roughly,
they correspond to Fig \ref{fig2} b) and c) of this work.
The droplet with the largest $AR$ detected in the experiment (1.99) correspond to a two-lobed shape, as shown in Fig. \ref{fig2} c).

As mentioned, so far there is no direct experimental information on the  angular momentum of the rotating droplets, and the rotational
axis has been determined only in a few cases. 
The shape of $^4$He droplets has been determined parametrizing them and  
computing the wide-angle diffraction patterns they produce, iteratively changing the parameters 
until matching the experimental diffraction patterns.\cite{Lan18} 
This procedure does not provide the angular momentum nor the direction of the
rotation axis, but supplies interesting information, 
in particular the  distance of the COM to the droplet 
surface along the axes used to describe the parametrized droplet surface. 
This is the rationale for displaying  $b_y^3/V$ {\it vs}.  the ratio $a_x/c_z$ in Fig. \ref{fig8}; 
when $a_x/c_z = 1$, $b_y^3/V = R^3/V = 3/(4 \pi)$, that can be used   to  compare 
the classical and DFT results. The dashed line shows the 
classical model result of Ref.  \onlinecite{Bal15}. The agreement between 
classical and DFT-TF calculations for $^3$He droplets
 is good, and even remarkable for prolate configurations. A similar good agreement was 
found between parametrized, vortex-hosting DFT, and classical $^4$He 
droplets.\cite{Anc18,Lan18} 

\section{Summary and outlook}

We have studied rotating $^3$He droplets combining a semiclassical 
Thomas-Fermi approach with the well established  
DFT formalism. 
We have shown that classical models for the equilibrium shapes of rotating drops which are 
subject to surface tension and centrifugal forces alone work remarkably well when they are applied to nanoscopic quantum
object as $^3$He droplets. Minor differences appear between their results  and the DFT ones that
are likely due to  a better description of the droplet surface and to quantum kinetic energy contributions
in the microscopic approach that, together with curvature energy and compressibility effects, are lacking in classical models. 

The DFT approach to rotating helium nanodroplets has been previously applied  to isotopically pure $^4$He superfluid droplets, 
allowing to clarify the influence of vortex arrays on their equilibrium shapes and
disclosing the presence of capillary waves and their interplay with vortices.\cite{Anc18,Oco19}
We have shown here that the DFT approach allows to describe as well
rotating quantum normal fluid $^3$He droplets  on a microscopic and firm basis. 
The availability of an accurate theoretical framework for studying spinning droplets 
of both isotopes is a crucial  ingredient necessary for addressing a 
far more challenging system, namely rotating mixed $^3$He-$^4$He 
droplets at very low temperatures, a work that is now in progress.

\begin{acknowledgments}
We thank Andrey Vilesov and Luis Egido  for useful discussions and exchanges. 
We are most indebted to Sam Butler for providing us with the results of the classical model calculations. This work has been 
performed under Grant No  FIS2017-87801-P (AEI/FEDER, UE).
M. B. thanks the Universit\'e F\'ed\'erale Toulouse Midi-Pyr\'en\'ees for financial support  throughout 
the ``Chaires d'Attractivit\'e 2014''  Programme IMDYNHE
\end{acknowledgments}

%\appendix*
%\setcounter{equation}{0}
%\section{Rotating TF model}

%\section*{Appendix}
%\setcounter{equation}{0}

\appendix*
\setcounter{equation}{0}
\section{}

We introduce in this Appendix the basics of the rotating TF model, as discussed in Ref. \onlinecite{Gra78}. 
In a rotating $^3$He droplet, the local momentum distribution of its atoms
 is altered due to the motion of the droplet as a whole. The Fermi sphere  is displaced from  $\mathbf{k}=0$ 
\begin{equation}
|\mathbf{k}(\mathbf{r}) - \mathbf{K}_R(\mathbf{r})| \leq k_F(\mathbf{r})
\label{A1}
\end{equation}
where $\hbar\mathbf{k}_F(\mathbf{r})$ is the Fermi momentum at point $\mathbf{r}$ and $\hbar\mathbf{K}_R(\mathbf{r})$ is the local
momentum due to rotation, namely
\begin{equation}
\mathbf{K}_R(\mathbf{r}) = \frac{m}{\hbar} \boldsymbol{\omega} \times \mathbf{r}
\label{A2}
\end{equation}
It is straightforward to obtain the expressions for quantities such as  the particle number $N$ 
\begin{equation}
N = \frac{2}{(2 \pi)^3} \int d \mathbf{r} \int_{|\mathbf{k}(\mathbf{r}) - \mathbf{K}_R(\mathbf{r})| \leq k_F(\mathbf{r})} d \mathbf{k}
= \frac{1}{3 \pi^2} \int d \mathbf{r} \, k_F^3(\mathbf{r}) 
\label{A3}
\end{equation} 
Thus the particle density is
\begin{equation}
\rho(\mathbf{r})=   \frac{k_F^3(\mathbf{r})}{3 \pi^2}  \Rightarrow  k_F(\mathbf{r}) = [3 \pi^2 \rho(\mathbf{r})]^{1/3}
\label{A4}
\end{equation} 
The kinetic energy is similarly obtained 
\begin{equation}
T = \frac{2}{(2 \pi)^3} \int d \mathbf{r} \int_{|\mathbf{k}(\mathbf{r}) - \mathbf{K}_R(\mathbf{r})| \leq k_F(\mathbf{r})}  d \mathbf{k} \,
\frac{\hbar^2 k^2}{2 m} 
\label{A5}
\end{equation} 
As done in Eq. (\ref{A3}) for $N$, if $\mathbf{k}(\mathbf{r}) - \mathbf{K}_R(\mathbf{r}) \equiv \mathbf{k}'(\mathbf{r})$, the Jacobian is $|J|=1$ and the kinetic energy 
becomes
\begin{equation} 
T = \frac{\hbar^2}{2 m} \int d \mathbf{r} \frac{2}{(2 \pi)^3}  \int_{|\mathbf{k}'(\mathbf{r})| \leq k_F(\mathbf{r})}  d \mathbf{k}'
[ k'^2 + K_R^2]
\label{A6}
\end{equation} 
(the integral of the cross term  $2 \, \mathbf{k}'(\mathbf{r}) \cdot \mathbf{K}_R(\mathbf{r})$ is zero). 
Hence,
\begin{equation} 
T = \frac{\hbar^2}{2 m} \int d \mathbf{r} \frac{2}{(2 \pi)^3} 
\left[ \frac{4 \pi}{5} k_F^5(\mathbf{r}) + K_R^2(\mathbf{r}) \frac{4 \pi}{3} k_F^3(\mathbf{r})\right]
\label{A7}
\end{equation}
and
\begin{equation} 
T = \frac{\hbar^2}{2 m} \int d \mathbf{r} \frac{3}{5} (3 \pi^2)^{2/3}\rho^{5/3}(\mathbf{r})   +
\frac{\hbar^2}{2 m} \int d \mathbf{r} \, K_R^2(\mathbf{r}) \, \rho(\mathbf{r}) 
\label{A8}
\end{equation}
The first term is the ordinary TF kinetic energy. The second term is easily identified with the rotation energy. If we take 
$\boldsymbol{\omega}$  in the $z$ direction $[\boldsymbol{\omega} =  \omega (0,0,1)]$ and substitute $\mathbf{K}_R(\mathbf{r})$ 
by its expression Eq. (\ref{A2}) one gets 
\begin{equation} 
\frac{\hbar^2}{2 m} \int d \mathbf{r} \, K_R^2(\mathbf{r})\,  \rho(\mathbf{r}) =
\frac{1}{2} m \omega^2 \int d \mathbf{r} (x^2+y^2) \rho(\mathbf{r}) \equiv \frac{1}{2} {\cal I} \omega^2
\label{A9}
\end{equation}
where we have introduced the definition of the moment of inertia ${\cal I}$ about the $z$ axis. It is worth seeing that  the TF
approximation leads naturally to a rigid-body rotation in the case of fermions. 

Let us calculate the angular momentum
\begin{eqnarray}
&&\mathbf{L} = \frac{2}{(2 \pi)^3} \hbar \int d \mathbf{r} \int_{|\mathbf{k}(\mathbf{r}) - \mathbf{K}_R(\mathbf{r})| \leq k_F(\mathbf{r})}
 d \mathbf{k} \;  (\mathbf{r} \times \mathbf{k})
 \\
\nonumber
&& 
= \frac{2}{(2 \pi)^3} \hbar \int d \mathbf{r} \int_{|\mathbf{k}'(\mathbf{r})| \leq k_F(\mathbf{r})}
 d \mathbf{k}' \; \mathbf{r} \times [\mathbf{k}'+ \mathbf{K}_R(\mathbf{r})]
 \\
\nonumber
&&
= \hbar  \int d \mathbf{r} [ \mathbf{r} \times  \mathbf{K}_R(\mathbf{r}) ] \rho(\mathbf{r}) =
m \int d \mathbf{r} [ \mathbf{r} \times  (\boldsymbol{\omega}\ \times \mathbf{r}) ] \rho(\mathbf{r})
\label{A10}
\end{eqnarray} 
In cartesian coordinates, 
\begin{equation} 
\mathbf{r} \times  (\boldsymbol{\omega}\ \times \mathbf{r}) = \omega [ -x z \hat{i} -y z \hat{j} +(x^2 + y^2) \hat{k}]
\label{A11}
\end{equation}
Thus, if  $\rho(x,y,z)$ is such that
\begin{equation} 
\rho(x,y,z) = \rho(-x,y,z) \quad {\rm and} \quad \rho(x,y,z) = \rho(x,-y,z)
\label{A12}
\end{equation}
or 
\begin{equation} 
\rho(x,y,z) = \rho(x,y,-z)
\label{A13}
\end{equation}
we get
\begin{equation} 
\mathbf{L} = m \boldsymbol{\omega}  \int d \mathbf{r} (x^2+y^2) \rho(\mathbf{r}) = {\cal I} \boldsymbol{\omega}
\label{A14}
\end{equation}
In their classical paper  
on drops under surface tension, Brown and Scriven\cite{Bro80}  have assumed that  droplets have reflective symmetry about their 
equator plane ($z=0$), and at least one meridional plane of reflective symmetry (either $x=0$ or $y=0$). 
 We have assumed that $z=0$ is a reflective plane of  symmetry and have taken Eq. (\ref{A14}) for the definition of  $\mathbf{L}$.


\begin{thebibliography}{99}
\bibitem{note}
We have analyzed  the density profiles along the coordinate axes for some  oblate and prolate 
configurations, and have found that surface diffuseness hardly depends on the deformation of the droplet.
\end{thebibliography}
\end{document}